\documentclass[sigconf]{acmart}
\AtBeginDocument{%
  }
\copyrightyear{2024} 
\acmYear{2024} 
\setcopyright{acmlicensed}\acmConference[ASE '24]{39th IEEE/ACM
International Conference on Automated Software Engineering }{October
27-November 1, 2024}{Sacramento, CA, USA}
\acmBooktitle{39th IEEE/ACM International Conference on Automated Software
Engineering (ASE '24), October 27-November 1, 2024, Sacramento, CA, USA}
\acmDOI{10.1145/3691620.3695351}
\acmISBN{979-8-4007-1248-7/24/10}

\usepackage{amsmath,amsfonts}
\usepackage{algorithmic}
\usepackage{graphicx}
\usepackage{textcomp}
\usepackage{xcolor}
\usepackage{hyperref}
\usepackage{subcaption}

\usepackage{algorithm}
\begin{document}

\title{DroneWiS: Automated Simulation Testing of small Unmanned Aerial Systems in Realistic Windy Conditions}

\author{Bohan Zhang}
\email{bohan.zhang.1@slu.edu}
\orcid{XX}
\affiliation{%
  \institution{Saint Louis University}
  \city{Saint Louis}
  \state{MO}
  \country{USA}
}

\author{Ankit Agrawal}
\email{ankit.agrawal.1@slu.edu}
\affiliation{%
  \institution{Saint Louis University}
  \city{Saint Louis}
  \state{MO}
  \country{USA}
}

\renewcommand{\shortauthors}{Zhang et al.}

\newcommand{\ankit}[1]{\textcolor{blue}{#1}}
\newcommand{\bohan}[1]{\textcolor{red}{#1}}

\newcommand{\DWFull}{Drone Wind Simulation}

\newcommand{\DW}{DroneWiS }
\newcommand{\dw}{DroneWiS}

\begin{abstract}
% \ankit{Rewrite the abstract - Focus on CFD Wind Simulation Not Google Tiles etc - those were the main contribution in the previous paper. }

% The continuous evolution of small Unmanned Aerial Systems (sUAS) demands advanced testing methodologies to ensure their safety and reliability. DRV 2.0, an upgrade of DroneReqValidator, introduces DroneWISP (Wind Simulation and Prediction), an automated feature that enhances simulation fidelity through dynamic wind conditions modeled with OpenFOAM\cite{jasak2009openfoam}. This innovation significantly improves the realism and accuracy of wind simulations across various environments, enabling more thorough testing of sUAS. By automating terrain data integration and streamlining setup, DroneWISP can be applied universally to any sUAS simulation platform. This advancement provides developers with a crucial tool to enhance the reliability and safety of aerial applications in complex environments. A comprehensive demonstration is available at \url{xxx}.

The continuous evolution of small Unmanned Aerial Systems (sUAS) demands advanced testing methodologies to ensure their safe and reliable operations in the real-world. To push the boundaries of sUAS simulation testing in realistic environments, we previously developed the DroneReqValidator (DRV) platform \cite{zhang2023dronereqvalidator}, allowing developers to automatically conduct simulation testing in digital twin of earth. In this paper, we present DRV 2.0, which introduces a novel component called DroneWiS (Drone Wind Simulation). DroneWiS allows sUAS developers to automatically simulate realistic windy conditions and test the resilience of sUAS against wind. Unlike current state-of-the-art simulation tools such as Gazebo and AirSim that only simulate basic wind conditions, DroneWiS leverages Computational Fluid Dynamics (CFD) to compute the unique wind flows caused by the interaction of wind with the objects in the environment such as buildings and uneven terrains. This simulation capability provides deeper insights to developers about the navigation capability of sUAS in challenging and realistic windy conditions. DroneWiS equips sUAS developers with a powerful tool to test, debug, and improve the reliability and safety of sUAS in real-world. A working demonstration is available at \url{https://youtu.be/khBHEBST8Wc}.

\end{abstract}



\maketitle

% \begin{IEEEkeywords}
% Simulation, Automated Analysis, Uncrewed Aerial Vehicles, Computational Fluid Dynamics
% \end{IEEEkeywords}

\section{Introduction}

The rapid advancements in small Uncrewed Aerial Systems (sUAS) has led to their increased use in various applications, including urban logistics and military operations in challenging terrains \cite{chen2022dronetalk, stodola2019cooperative}. However, ensuring the safety and reliability of these systems in adverse environmental conditions, such as complex wind dynamics remains a significant challenge \cite{ilachinski2017artificial}. When testing sUAS under windy conditions, sUAS developers suffer from two primary challenges.

Traditional sUAS simulation tools, such as Gazebo and AirSim, use oversimplified wind models that ignore complex interactions with terrain and buildings. These models typically feature stochastic wind fluctuations or constant wind velocity, resulting in inaccurate simulations.
Such oversights are particularly concerning because wind patterns change drastically near uneven terrain and buildings, posing significant safety risks to sUAS due to discrepancies between simulated and real-world wind flows.
Additionally, computing wind flows using Computational Fluid Dynamics (CFD) in a 3D environment is a manual and labor-intensive process, involving a 3D scene designer to create the environment and a CFD engineer to compute wind flows using tools like OpenFoam \cite{jasak2009openfoam}. This process must be repeated for each unique environment, highlighting the need for automation to allow sUAS developers to create complex wind patterns more efficiently.

These two primary limitations of sUAS simulation tools hinder developers from testing their systems in realistic wind conditions, resulting in a significant gap between simulation results and real-world sUAS operations. To mitigate the limitations of existing sUAS simulation tools, we introduce \DW (\DWFull), a novel module integrated into our established platform, DroneReqValidator(DRV) \cite{zhang2023dronereqvalidator, agrawal2023requirements}. This enhancement evolves DRV into its next generation, DRV 2.0.

The \DW module uses state-of-the-art CFD software, OpenFOAM \cite{jasak2009openfoam}, with a terrain scanning algorithm in Unreal Engine, to automatically compute and simulate desired wind conditions in any given 3D environment. This allows sUAS developers to examine the effect of realistic wind flow around buildings and complex structures on sUAS trajectory. This innovative approach generates high-fidelity, dynamic wind vector data for various wind types, including uniform, turbulent, and multi-source winds.

Next, we discuss the architectural design of \DW (Section \ref{sec:arch}) describing the algorithms employed to automatically compute CFD-based wind flows within a given 3D environment (Algorithm \ref{algo1}). Furthermore, we also describe the integration of \DW within DRV for evaluation purposes and demonstrate the platform's efficacy in sUAS testing (Section \ref{sec:senario}).

% DroneReqValidator (DRV)\cite{zhang2023dronereqvalidator}, initially developed as a comprehensive simulation ecosystem for sUAS, aimed to bridge the gap between software engineering and realistic 3D environment creation. It significantly reduced the time and resources required for manual environment setups. By integrating Unreal Engine, Google Earth digital twin models, and AirSim APIs, DRV facilitated the automatic generation of realistic simulation environments, enabling the simulation of multiple UAVs under varied testing conditions \cite{unrealengine, googlemaps2021tile, shah2018airsim}.
\section{System Design}

\label{sec:arch}

\begin{figure*}[htbp]
    \centering
    \begin{subfigure}{0.28\textwidth}
        \centering
        \includegraphics[width=\textwidth]{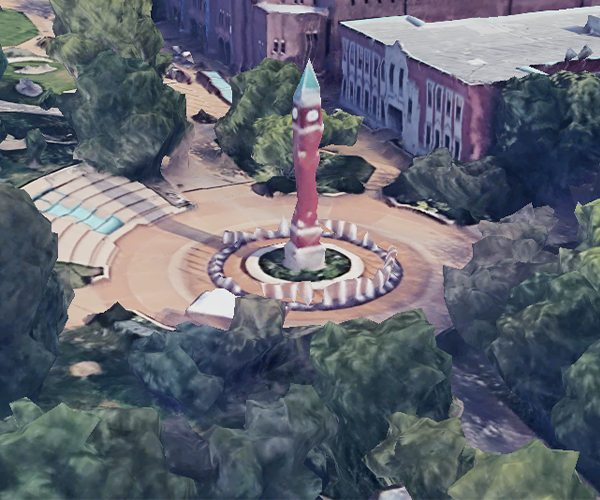}
        \caption{Google Earth Digital Twin Model}
        \label{fig:screenshot1}
    \end{subfigure}
    \hfill
    \begin{subfigure}{0.28\textwidth}
        \centering
        \includegraphics[width=\textwidth]{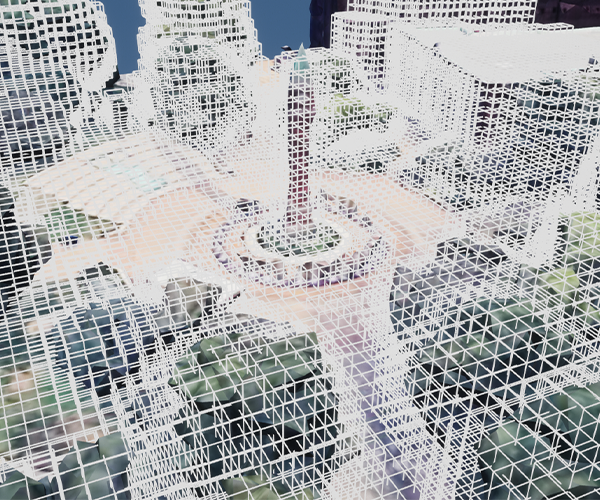}
        \caption{Scanned Voxel Grid}
        \label{fig:screenshot2}
    \end{subfigure}
    \hfill
    \begin{subfigure}{0.28\textwidth}
        \centering
        \includegraphics[width=\textwidth]{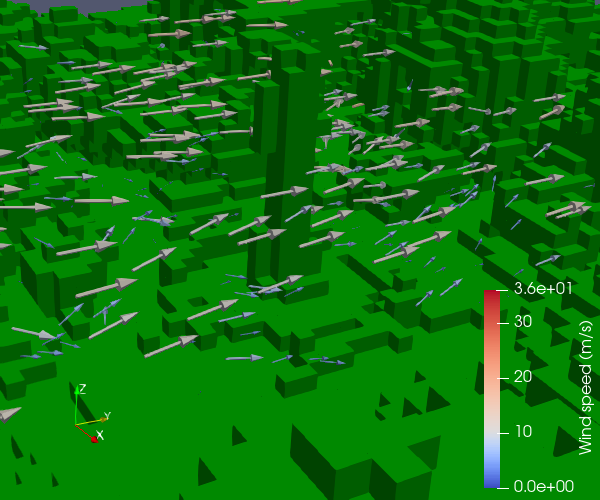}
        \caption{OpenFOAM Wind-Flow Computation}
        \label{fig:screenshot3}
    \end{subfigure}
    \vspace{-1em} % Adjust this value to reduce vertical spacing
    \caption{
    The figure shows three views of the same environment: (a) the initial terrain, (b) the voxel grid from the Terrain Scanning Algorithm, and (c) wind simulation results from OpenFOAM, with arrows indicating wind velocity and direction.
    }
    \label{fig:screenshots}
    \vspace{-1em}
\end{figure*}

\begin{figure}[htbp]
    \centering
    \includegraphics[width=0.95\columnwidth]{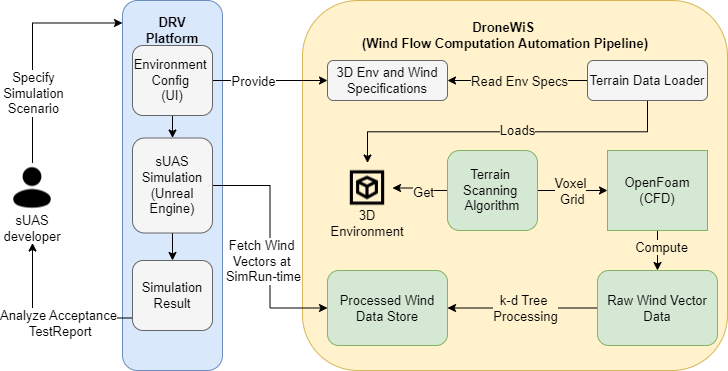}
    \caption{\DW Architecture and Integration with DRV}
    \label{fig:wisp_arch}
    \vspace{-3pt}
\end{figure}

% \subsection{Background}

\noindent \textbf{DRV Platform Background:} In our previous work, we presented the DRV platform \cite{zhang2023dronereqvalidator, agrawal2023requirements} which uses a modular client-server architecture, comprising a React-based front-end and a Python/Flask back-end. The user-friendly interface allows developers to configure realistic simulation environments, including specific parts of the Google Digital Twin model where they wish to simulate one or more UAVs. The back-end automatically retrieves 3D environment data from the Google Earth 3D API to create the simulation. It uses AirSim APIs and Unreal Engine to simulate the behavior of single or multiple sUAS in the user-specified geolocations.

\noindent\textbf{\DW}: 
Figure \ref{fig:wisp_arch} presents the overall architecture of \DW and its integration with the DRV platform. \DW receives inputs from the DRV User Interface regarding the simulation region, which corresponds to real-world geo-coordinates (e.g., Latitude and longitude of Saint Louis University), and the wind conditions developers wish to simulate. In addition to normal and turbulent wind simulations, \DW allows users to specify multi-source winds. For example, it can simulate normal wind from the east of Chicago at 10 meters per second and turbulent wind from the west at 30 meters per second with 40\% fluctuation, exposing sUAS to complex urban wind conditions. 

Based on these inputs, \DW automatically produces the wind flow data. The automation pipeline  includes two primary components: 1) a terrain scanning algorithm to understand the details of obstacles and elements in the environment, 2) OpenFoam, a CFD tool,  to automatically compute the wind vectors in the environment, and process them for quick read operation during simulation run-time.

\subsection{Terrain Scanning}
\label{sec:terrainScan}
To model wind flow around complex real-world objects (such as buildings), we need to represent the digital twin model of the environment in a format that can be easily processed by computational algorithms. This includes identifying which parts of the 3D environment are occupied by objects and which parts are empty where wind flow needs to be computed. Therefore, we developed a terrain scanning algorithm (Algorithm \ref{algo1}) to efficiently scan the 3D digital twin model (Fig. \ref{fig:screenshot1}). 

The algorithm takes the 3D environment as input and overlays a Volumetric Pixel Grid (Voxel Grid shown in Fig \ref{fig:screenshot2}), where each cell (a 1-meter cube) acts as a sampling point (Algo 1 - Lines 1-3). For each cell in the grid, we need to identify if it is occupied by an object or empty. Thus, for each cell (Line 4), we run the function ObstructionCheck, which takes the center of the grid cell as input (Line 5) and returns false if the cell is empty or true if it is occupied (Line 6).

\setlength{\textfloatsep}{2pt}
\begin{algorithm}
\caption{Scanning Algorithm}
\label{algo1}
\begin{algorithmic}[1]

\STATE \textbf{Input:} $scanRange$ \COMMENT{$x \cdot y \cdot z$}
\STATE \textbf{Output:} $voxelGrid$

\STATE $voxelGrid \gets \{\}$

\FOR{each $coordinate$ in $scanRange$}
    \STATE $cellCenter \gets$ center of the current $coordinate$ cell
    \STATE $hit \gets$ ObstructionCheck($cellCenter$)
    \IF{$hit$}
        \STATE add $coordinate$ to $voxelGrid$
    \ENDIF
\ENDFOR
\RETURN $voxelGrid$
\STATE \textbf{function ObstructionCheck($cellCenter$)}
    \STATE \quad \textbf{Input:} $cellCenter$
    \STATE \quad \textbf{Output:} $bool$
    \STATE \quad $faces \gets$ Generate 6 faces of the voxel centered at $cellCenter$
    \FOR{each pair of opposite face centers in $faces$}
        \STATE \quad $hasLineOfSight \gets$ lineTrace{(face1, face2)}
        \IF{Not $hasLineOfSight$}
           \RETURN true
        \ENDIF
    \ENDFOR
    \quad \RETURN false
\STATE \textbf{end function}
\end{algorithmic}
\end{algorithm}

The ObstructionCheck function (Lines 13-26) uses Unreal Engine ray tracing to determine if a cell is empty or occupied. For example, consider a cube cell in the volumetric grid; this cell has six faces, each with a central point denoted as $P_1$ to $P_6$. Here, $P_1$ to $P_4$ correspond to the central point of the back, right, front, and left sides of the cube, while $P_5$ and $P_6$ are the top and bottom sides (Line 16). We have three unique pairs of opposite faces: $P_{1,3}$, $P_{2,4}$, and $P_{5,6}$. From each pair (Line 17), we trace two rays, resulting in a total of six rays. Algorithm \ref{algo1} adds a cell to the resulting voxelGrid if there is a visibility obstruction between any pair of opposite faces, indicating the presence of an obstacle. The remaining cells represent open spaces. The obstructed cells are represented by green color in Figure \ref{fig:screenshot3}.

\subsection{Wind Flow Computation and Processing}
After \DW scans the environment, the resulting voxel grid is transferred to OpenFoam for processing voxel grids and computing wind vectors. \DW automatically configures OpenFoam according to user-specified wind conditions, which include wind type (normal, turbulent, multi-source), wind direction, and wind velocity. Therefore, OpenFoam takes the voxel grid and wind characteristics as inputs to compute wind flows.

After OpenFoam finishes computing the wind flow, it generates wind vectors that describe the velocity and direction of wind for each empty grid cell. We utilize an advanced k-d Tree preprocessing algorithm \cite{kdtreeskrodzki2019kd} to optimize random access speed of data using nearest neighbor queries and store the constructed k-d Tree in the wind speed database. This processing step is crucial to retrieve the wind vector from the database during simulation run-time in negligible time and ensure that wind forces are applied to the simulated UAV for each time step of the simulation.

During the simulation, DRV retrieves the wind vector at the current geographical coordinates of the simulated sUAS from \DW and applies the wind forces to impact the flight path of the UAV based on the computed wind at that geo-location. As shown in the architectural diagram in Figure \ref{fig:wisp_arch}, this design completely separates the responsibilities of wind computation and corresponding sUAS simulation. Therefore, this design allows existing sUAS simulation platforms such as Gazebo to also utilize \DW to access wind data at specific sUAS locations, apply wind forces, and simulate sUAS behavior in variety of wind conditions.

\subsection{Automation Computational Analysis}
\begin{figure}[htbp]
   \centering
   \includegraphics[width=0.85\columnwidth]
   {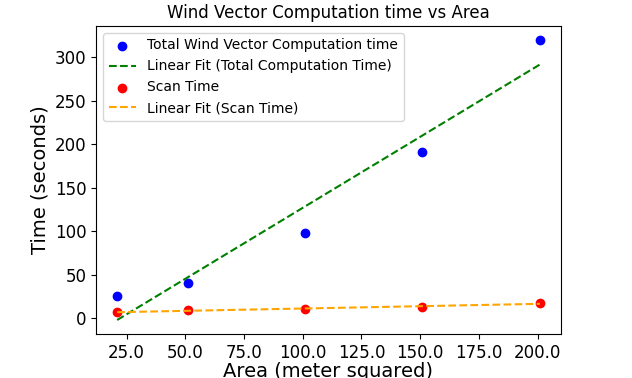}
   \caption{Terrain Scan and Wind Flow Computation Time}
   \label{fig: precision vs. cfd time graph}
   \vspace{-2em}
\end{figure}

To assess the efficiency of the scanning algorithm, we analyzed its performance across various scanning ranges, as depicted in Fig. \ref{fig: precision vs. cfd time graph}. The graph displays two lines: one for scanning time and another for wind computation and processing time across 3D environment of various sizes. We can observe that the terrain scanning algorithm operates quickly and efficiently without significantly impacting the overall computation time, unlike the time-intensive calculations of OpenFoam CFD for wind computations. Since CFD methods are traditionally computationally intensive, there is a growing interest in transitioning towards predictive approaches for wind flow analysis. As part of our future plans, we aim to optimize wind flow computation time by leveraging physics-informed neural networks. 

\section{Application of \DW in sUAS Testing}

\label{sec:senario}
To rigorously assess the accuracy of wind computation and the responsiveness of sUAS to wind conditions in simulations, we conducted a comparative analysis with Airsim \cite{shah2018airsim}. Our primary objective was to evaluate how wind conditions influenced sUAS flight paths and to determine the extent to which their behavior appeared ``organic''—that is, realistically unpredictable and natural—under similar conditions across different simulation platforms. We used scenario-based approach to validate the effectiveness of \DW in simulation testing\cite{ryser1999scenario}. 

% Scenarios were selected from real-world sUAS incidents and hazards caused by unexpected wind conditions. Our goal was to evaluate whether \DW could replicate or identify these wind-induced issues. 

\textbf{Environment Set-Up:} We selected several points of interest on the Saint Louis University campus and used the DRV front-end to specify these simulation regions along with the dynamic wind conditions that an sUAS developer would want to simulate. Specifically, we set gusty wind flow at 10 meters per second from various directions with a 30 percent fluctuation in velocity. These inputs were then sent to \DW, which scanned the environment using Algorithm 1 and produced the processed wind vector database that DRV used to apply wind forces during simulation run-time.

\begin{figure}[htbp]
    \centering
    \includegraphics[width=0.9\columnwidth]{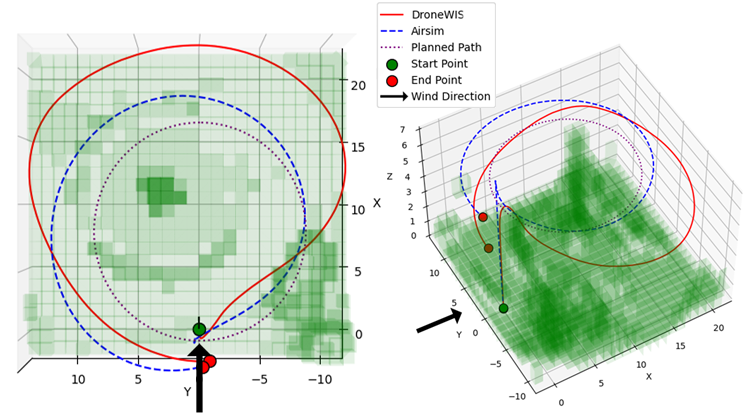}\vspace{-1em}
    \caption{Scenario - Circular flight around stationary object}
    \label{fig:rq3p1}
    \vspace{-1em}
\end{figure}

\noindent{\textbf{Scenario 1: Circular Flight Path in Gusty Winds}}
The objective of this scenario was to evaluate the capability of DroneWIS in simulating dynamic wind behavior in an urban environment and to observe its impact on sUAS flight paths. In this scenario, we utilized the digital twin environment as depicted in Fig. \ref{fig:screenshot1}, and used \DW to generate the Voxel Grid as presented in \ref{fig:screenshot3}. We also used AirSim to simulate sUAS in the same environment and employed its default wind model to observe the flight path for comparative analysis.

\noindent\textbf{Analysis}: The observed flight paths is presented in the Figure \ref{fig:rq3p1}. As illustrated, the circular flight path of the sUAS using \DW (red line) exhibits more dynamic changes and variance in wind speed around the stationary tower in the center. The wind flow diverges after hitting the tower, causing wider flight deviations on the other side. In contrast, the flight path using AirSim (blue dashed line) does not show such dynamic changes because it assumes that wind flows uniformly and does not interact with environmental elements. This results in a more linear and less responsive flight path. The comparison highlights \dw's capability to account for the impact of environmental elements, such as buildings and other obstacles, to generate realistic wind patterns and observe how these wind flows affect the sUAS flight path.

% This comparison highlights the potential impact of \DW on flight performance, offering a significant departure from the predictable uniform wind model towards a more realistic and challenging flight environment. Our observations reveal that under \DW, the UAV must constantly adjust its flight dynamics to compensate for the fluctuating wind conditions, providing a robust testbed for evaluating UAV resilience and adaptive control algorithms.

% As can be observed from Figure \ref{fig:rq3p1}, wind velocity is relatively high above the shorter building, which then rebounded after colliding with the taller building to create a whirlwind. Due to this, the sUAS had trouble taking off, as can be seen from its irregular flight path near the take-off point. When the sUAS reached the altitude of the shorter building, high wind forces pushed it downwards in the whirlwind area, resulting in a crash. 

% In contrast, Using AirSim's default uniform wind model, the simulation was successful, and no significant drift in the fight path was observed when compared to the planned flight path. This highlights the applicability of \DW in simulation testing and enabling sUAS developers to discover the limits of their sUAS applications in specific scenarios before field testing or deployment. 

\noindent{\textbf{Scenario 2: Unexpected Wind during Takeoff}}

We recreated a 2022 FAA-reported incident where a drone crashed while taking off between two buildings (one taller than the other) to capture target imagery. Although the buildings shielded the takeoff location from the wind’s direct effects, as soon as the sUAS reached a higher altitude, high wind velocity blew the UAV off course.

As part of our evaluation, we recreated this scenario to verify if \DW can detect such deviations. We used DRV to pick an environment for sUAS simulation where two buildings of different heights are close to each other, configured to simulate wind speed of 20 meters per second blowing towards the north. \DW uses this as input to compute the wind vectors in that environment. 

\begin{figure}[t]
    \centering
    \includegraphics[width=0.9\columnwidth]{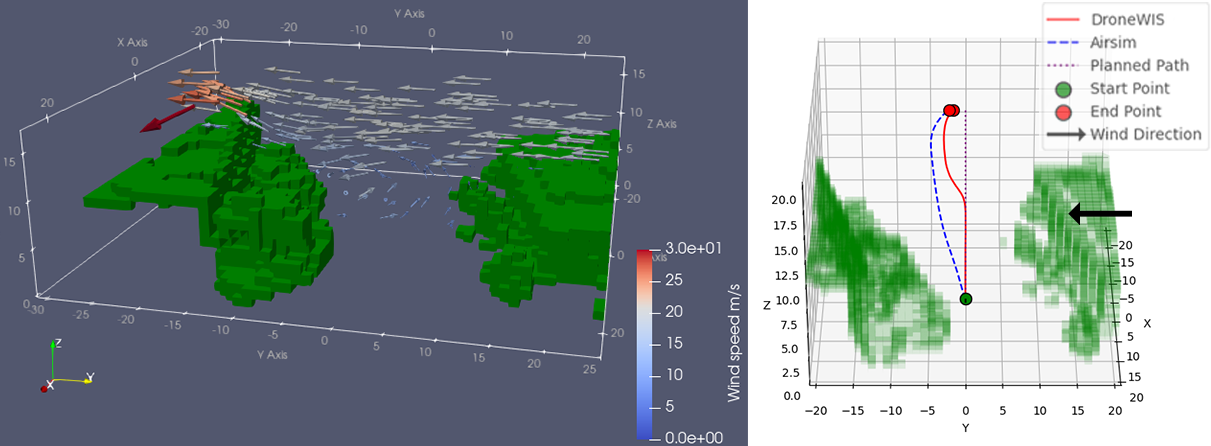}
    \caption{Scenario - Unexpected wind during takeoff}
    \label{fig:rq3p2}
\end{figure}

\noindent\textbf{Analysis:} As shown in Figure \ref{fig:rq3p2}, the wind velocity above the buildings is notably higher, forming an area with increased wind speeds compared to lower altitudes. Initially, the sUAS encountered minimal difficulty during takeoff, as evidenced by its flight path near the takeoff point. However, as the sUAS ascended to an altitude parallel to the rooftops of the buildings, strong wind forces pushed it in the direction of the wind, causing it to fly chaotically.

In contrast, when using AirSim's wind simulations, the simulated flight path showed a smooth curve. This indicates that wind effects on sUAS were present even at low altitudes, where buildings obstruct the wind coming from the source direction. This smooth flight path occurs because AirSim's wind simulations do not account for obstacles such as buildings in their computation and simulation of wind flows.

% \DW in simulation testing, allowing sUAS developers to identify the limits of their applications in specific scenarios before field testing or deployment.

% For this scenario, we selected a commercial application of sUAS, specifically roof inspection \cite{DroneRoo48:online, EagleVie56:online, hezaveh2017roof}. During simulation testing, we aimed to analyze how complex wind conditions during a roof inspection could affect sUAS behavior. Using Unreal Engine, we designed a 3D structure resembling a typical residential house with a sloped roof. We defined the sUAS flight path to closely inspect every part of the roof. We modeled a hypothetical gusty cross-wind blowing in the +x and -y directions to simulate the unpredictable wind conditions commonly encountered in real-world scenarios. The environment, wind vectors, wind flow direction, and the sUAS planned and observed flight paths are shown in \ref{fig:rq3p2}.

% After takeoff, the sUAS followed its intended path along the roof's edge with minimal drift, demonstrating stability under normal wind conditions (velocity below 10 m/s). However, upon reaching the roof's peak at point G in Figure \ref{fig:rq3p2}, it encountered significantly higher wind velocities. This increase was due to its positioning between two wind sources and wind acceleration from the roof's slope. As a result, the sUAS struggled to compensate for the intensified wind and began drifting in the final wind direction while descending. This ultimately led to a collision with the ground at point H in Figure.\ref{fig:rq3p2}. 

\noindent\textbf{Summary of our Evaluation} - Using \DW, we replicated the system failures due to wind that have been reported in past incidents, and generated insights regarding system robustness in hypothetical windy conditions. While these are simulation results and require further validation under real-world windy conditions, the insights generated by \DW are invaluable resources for sUAS developers during the initial development stages. Additionally, systematic simulation testing under diverse wind conditions can support safety analysis processing by using simulation artifacts as evidence to claim/refute safety and can also be automated to integrate automatically into Safety Assurance Cases\cite{agrawal2019leveraging, agrawal2023leveraging}. 

% Moreover, when simulation testing using \DW is systematically conducted under a diverse range of wind conditions, the results can serve as evidences to support arguments about the safety of the sUAS in unexpected windy situations and can also be automatically augmented to Safety Assurance Cases .

\section{Conclusion}
\label{sec:conclusion}
We introduced \dw, a novel component, integrated with the original DRV platform, that enables sUAS developers to conduct simulation testing of sUAS against realistic, complex, and CFD-driven wind flows in a 3D environment. Our comparative analysis of \DW with AirSim's wind simulations demonstrated significant improvements in wind effect fidelity, offering sUAS developers a powerful means to test and ensure safe operation of sUAS in real-world. In future work, we plan to adopt a machine learning-based approach to enhance the efficiency of wind vector generation for large-scale environments, support a broader range of wind conditions, and improve the fidelity of wind forces applied to sUAS.

\section{CODE AVAILABILITY}
\label{sec:code}

Code-base of DRV and \DW is available on our public GitHub repository:\url{https://github.com/UAVLab-SLU/DRV_public}.

\section{Acknowledgements}
The work in this paper was
funded under USA National Aeronautics and Space Administration (NASA) Grant Number: 80NSSC23M0058

\bibliographystyle{ACM-Reference-Format}
\bibliography{software, wind,droneSim}


\begin{thebibliography}{11}

%%% the Web address.

\ifx \showCODEN    \undefined \def \showCODEN     #1{\unskip}     \fi
\ifx \showDOI      \undefined \def \showDOI       #1{#1}\fi
\ifx \showISBNx    \undefined \def \showISBNx     #1{\unskip}     \fi
\ifx \showISBNxiii \undefined \def \showISBNxiii  #1{\unskip}     \fi
\ifx \showISSN     \undefined \def \showISSN      #1{\unskip}     \fi
\ifx \showLCCN     \undefined \def \showLCCN      #1{\unskip}     \fi
\ifx \shownote     \undefined \def \shownote      #1{#1}          \fi
\ifx \showarticletitle \undefined \def \showarticletitle #1{#1}   \fi
\ifx \showURL      \undefined \def \showURL       {\relax}        \fi
% invisible to TeX
\providecommand\bibfield[2]{#2}
\providecommand\bibinfo[2]{#2}
\providecommand\natexlab[1]{#1}
\providecommand\showeprint[2][]{arXiv:#2}

\bibitem[Agrawal and Cleland-Huang(2023)]%
        {agrawal2023leveraging}
\bibfield{author}{\bibinfo{person}{Ankit Agrawal} {and} \bibinfo{person}{Jane Cleland-Huang}.} \bibinfo{year}{2023}\natexlab{}.
\newblock \showarticletitle{Leveraging Traceability to Integrate Safety Analysis Artifacts into the Software Development Process}. In \bibinfo{booktitle}{\emph{2023 IEEE 31st International Requirements Engineering Conference Workshops (REW)}}. IEEE, \bibinfo{pages}{475--478}.
\newblock


\bibitem[Agrawal et~al\mbox{.}(2019)]%
        {agrawal2019leveraging}
\bibfield{author}{\bibinfo{person}{Ankit Agrawal}, \bibinfo{person}{Seyedehzahra Khoshmanesh}, \bibinfo{person}{Michael Vierhauser}, \bibinfo{person}{Mona Rahimi}, \bibinfo{person}{Jane Cleland-Huang}, {and} \bibinfo{person}{Robyn Lutz}.} \bibinfo{year}{2019}\natexlab{}.
\newblock \showarticletitle{Leveraging artifact trees to evolve and reuse safety cases}. In \bibinfo{booktitle}{\emph{2019 IEEE/ACM 41st International Conference on Software Engineering (ICSE)}}. IEEE, \bibinfo{pages}{1222--1233}.
\newblock


\bibitem[Agrawal et~al\mbox{.}(2023)]%
        {agrawal2023requirements}
\bibfield{author}{\bibinfo{person}{Ankit Agrawal}, \bibinfo{person}{Bohan Zhang}, \bibinfo{person}{Yashaswini Shivalingaiah}, \bibinfo{person}{Michael Vierhauser}, {and} \bibinfo{person}{Jane Cleland-Huang}.} \bibinfo{year}{2023}\natexlab{}.
\newblock \showarticletitle{A Requirements-Driven Platform for Validating Field Operations of Small Uncrewed Aerial Vehicles}. In \bibinfo{booktitle}{\emph{2023 IEEE 31st International Requirements Engineering Conference (RE)}}. IEEE, \bibinfo{pages}{29--40}.
\newblock


\bibitem[Chen et~al\mbox{.}(2022)]%
        {chen2022dronetalk}
\bibfield{author}{\bibinfo{person}{Kuan-Wen Chen}, \bibinfo{person}{Ming-Ru Xie}, \bibinfo{person}{Yu-Min Chen}, \bibinfo{person}{Ting-Tsan Chu}, {and} \bibinfo{person}{Yi-Bing Lin}.} \bibinfo{year}{2022}\natexlab{}.
\newblock \showarticletitle{DroneTalk: An Internet-of-Things-based drone system for last-mile drone delivery}.
\newblock \bibinfo{journal}{\emph{IEEE Transactions on Intelligent Transportation Systems}} \bibinfo{volume}{23}, \bibinfo{number}{9} (\bibinfo{year}{2022}), \bibinfo{pages}{15204--15217}.
\newblock


\bibitem[Ilachinski(2017)]%
        {ilachinski2017artificial}
\bibfield{author}{\bibinfo{person}{Andrew Ilachinski}.} \bibinfo{year}{2017}\natexlab{}.
\newblock \showarticletitle{Artificial Intelligence and Autonomy: Opportunities and Challenges}.
\newblock  (\bibinfo{year}{2017}).
\newblock


\bibitem[Jasak(2009)]%
        {jasak2009openfoam}
\bibfield{author}{\bibinfo{person}{Hrvoje Jasak}.} \bibinfo{year}{2009}\natexlab{}.
\newblock \showarticletitle{OpenFOAM: open source CFD in research and industry}.
\newblock \bibinfo{journal}{\emph{International Journal of Naval Architecture and Ocean Engineering}} \bibinfo{volume}{1}, \bibinfo{number}{2} (\bibinfo{year}{2009}), \bibinfo{pages}{89--94}.
\newblock


\bibitem[Ryser and Glinz(1999)]%
        {ryser1999scenario}
\bibfield{author}{\bibinfo{person}{Johannes Ryser} {and} \bibinfo{person}{Martin Glinz}.} \bibinfo{year}{1999}\natexlab{}.
\newblock \showarticletitle{A scenario-based approach to validating and testing software systems using statecharts}.
\newblock  (\bibinfo{year}{1999}).
\newblock


\bibitem[Shah et~al\mbox{.}(2018)]%
        {shah2018airsim}
\bibfield{author}{\bibinfo{person}{Shital Shah}, \bibinfo{person}{Debadeepta Dey}, \bibinfo{person}{Chris Lovett}, {and} \bibinfo{person}{Ashish Kapoor}.} \bibinfo{year}{2018}\natexlab{}.
\newblock \showarticletitle{Airsim: High-fidelity visual and physical simulation for autonomous vehicles}. In \bibinfo{booktitle}{\emph{Field and service robotics}}. Springer, \bibinfo{pages}{621--635}.
\newblock


\bibitem[Skrodzki(2019)]%
        {kdtreeskrodzki2019kd}
\bibfield{author}{\bibinfo{person}{Martin Skrodzki}.} \bibinfo{year}{2019}\natexlab{}.
\newblock \showarticletitle{The kd tree data structure and a proof for neighborhood computation in expected logarithmic time}.
\newblock \bibinfo{journal}{\emph{arXiv preprint arXiv:1903.04936}} (\bibinfo{year}{2019}).
\newblock


\bibitem[Stodola et~al\mbox{.}(2019)]%
        {stodola2019cooperative}
\bibfield{author}{\bibinfo{person}{Petr Stodola}, \bibinfo{person}{Jan Drozd}, \bibinfo{person}{Jan Mazal}, \bibinfo{person}{Jan Hodick{\`y}}, {and} \bibinfo{person}{Dalibor Proch{\'a}zka}.} \bibinfo{year}{2019}\natexlab{}.
\newblock \showarticletitle{Cooperative unmanned aerial system reconnaissance in a complex urban environment and uneven terrain}.
\newblock \bibinfo{journal}{\emph{Sensors}} \bibinfo{volume}{19}, \bibinfo{number}{17} (\bibinfo{year}{2019}), \bibinfo{pages}{3754}.
\newblock


\bibitem[Zhang et~al\mbox{.}(2023)]%
        {zhang2023dronereqvalidator}
\bibfield{author}{\bibinfo{person}{Bohan Zhang}, \bibinfo{person}{Yashaswini Shivalingaiah}, {and} \bibinfo{person}{Ankit Agrawal}.} \bibinfo{year}{2023}\natexlab{}.
\newblock \showarticletitle{DroneReqValidator: Facilitating High Fidelity Simulation Testing for Uncrewed Aerial Systems Developers}. In \bibinfo{booktitle}{\emph{2023 38th IEEE/ACM International Conference on Automated Software Engineering (ASE)}}. IEEE, \bibinfo{pages}{2082--2085}.
\newblock


\end{thebibliography}
% \bibliography{software}

\end{document}